\documentclass[letter, 12pt]{article}
\usepackage{graphicx} % Required for inserting images
\usepackage[T1]{fontenc}
\usepackage[english]{babel}
\usepackage{amsmath, amssymb}
\usepackage[table]{xcolor}
\usepackage{times}
\usepackage{helvet}
\usepackage{sectsty} % use different fonts for different sections
\allsectionsfont{\sffamily} % for sections use sans serif
\usepackage[labelfont={bf,sf}]{caption} % customize captions
\usepackage{booktabs} % nicer tables
\usepackage{hyperref} % URLs and other refs
\usepackage{multirow}
\usepackage{pdflscape} % rotated landscape
\usepackage{afterpage} % floating landscape
\usepackage[doublespacing]{setspace}
\usepackage{orcidlink} % for ORCID symbol with link
\usepackage{bbold}
\usepackage{enumitem}
\usepackage{natbib}
\usepackage{doi}
\usepackage{tikz}
\usepackage{enumitem}
\usepackage{comment}
\usetikzlibrary{positioning, arrows, fadings,decorations.pathmorphing,arrows.meta, shapes.arrows}
\tikzset{>=stealth'} % arrow style
\usepackage{longtable}
\definecolor{pigment}{rgb}{0.2, 0.2, 0.6}
\definecolor{lightgray}{gray}{0.9}
\hypersetup{
  colorlinks = true,    %Colours links instead of ugly boxes
  urlcolor   = pigment, %Colour for external hyperlinks
  linkcolor  = black,   %Colour of internal links
  citecolor  = pigment  %Colour of citations
}

\usepackage{geometry}
\newif\ifanonymize % Boolean variable to control anonymization
\newcommand{\anonymize}[1]{%
  \ifanonymize
    \phantom{#1}%
  \else
    #1%
  \fi
}
\anonymizefalse % set to false to deanonymize
\title{\vspace{-4em}
\textsf{\textbf{Living Synthetic Benchmarks:\\A Neutral and Cumulative Framework for Simulation Studies}
}}

\usepackage{authblk}
\author[1]{
\anonymize{František~Bartoš \orcidlink{0000-0002-0018-5573}}
}
\author[2]{
\anonymize{Samuel~Pawel \orcidlink{0000-0003-2779-320X}}
}

\author[3]{
\anonymize{Björn~S.~Siepe \orcidlink{0000-0002-9558-4648}}
}
\affil[1]{
  \anonymize{
  Department of Psychological Methods,
  University of Amsterdam 
  }
}
\affil[2]{
  \anonymize{
  Epidemiology, Biostatistics and Prevention Institute,
  Center for Reproducible Science and Research Synthesis,
  University of Zurich
  }
}
\affil[3]{
  \anonymize{
  Psychological Methods Lab, 
  Department of Psychology, 
  University of Marburg
  }
}

\begin{document}

\begin{onehalfspacing}
\maketitle
\end{onehalfspacing}

\begin{abstract}
\noindent
Simulation studies are widely used to evaluate statistical methods. However, new methods are often introduced and evaluated using data-generating mechanisms (DGMs) devised by the same authors. This coupling creates misaligned incentives, e.g., the need to demonstrate the superiority of new methods, potentially compromising the neutrality of simulation studies. Furthermore, results of simulation studies are often difficult to compare due to differences in DGMs, competing methods, and performance measures. This fragmentation can lead to conflicting conclusions, hinder methodological progress, and delay the adoption of effective methods. To address these challenges, we introduce the concept of \emph{living synthetic benchmarks}. The key idea is to disentangle method and simulation study development and continuously update the benchmark whenever a new DGM, method, or performance measure becomes available. This separation benefits the neutrality of method evaluation, emphasizes the development of both methods and DGMs, and enables systematic comparisons. In this paper, we outline a blueprint for building and maintaining such benchmarks, discuss the technical and organizational challenges of implementation, and demonstrate feasibility with a prototype benchmark for publication bias adjustment methods. We conclude that living synthetic benchmarks have the potential to foster neutral, reproducible, and cumulative evaluation of methods, benefiting both method developers and users.
\end{abstract}

\section{Introduction}
\begin{quote}
    ``\emph{In fact it is very difficult to run an honest simulation comparison, and easy to inadvertently cheat by choosing favorable examples, or by not putting as much effort into optimizing the dull old standard as the exciting new challenger.}''
\end{quote}
\begin{flushright} Bradley Efron commenting on Breiman \cite[p. 219]{breiman2001} \end{flushright}

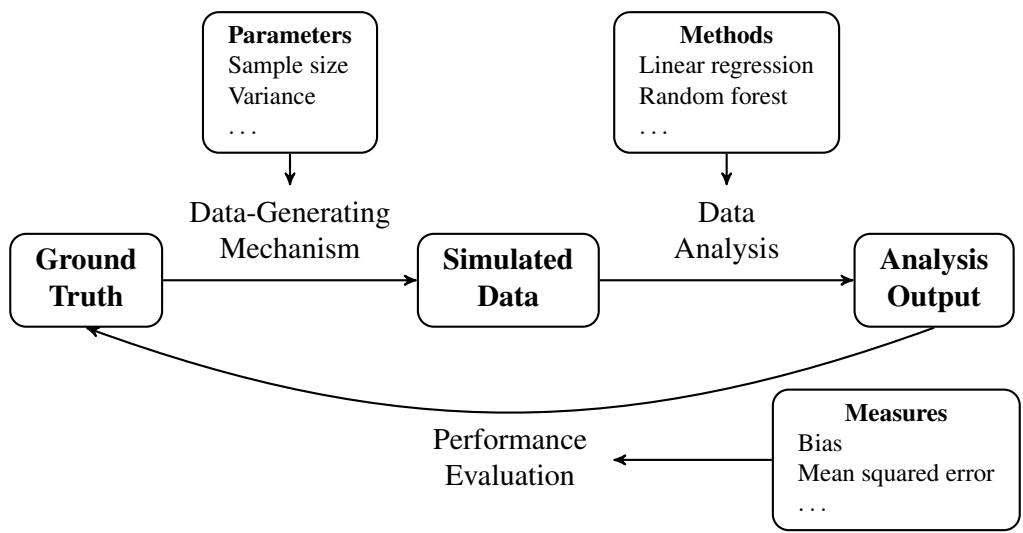
\begin{figure*}
    \centering
\begin{tikzpicture}[thick,every node/.style={scale=0.95}]

    % nodes
    \node [rectangle, draw, rounded corners = 0.5em] (truth)
    {\begin{tabular}{c} \textbf{Ground} \\
       \textbf{Truth}
     \end{tabular}};

   \node [rectangle, draw, rounded corners = 0.5em] (simdat) [right = 8em of truth]
   {\begin{tabular}{c} \textbf{Simulated} \\
      \textbf{Data}
    \end{tabular}};

  \node [rectangle, draw, rounded corners = 0.5em] (output) [right = 8em of simdat]
  {\begin{tabular}{c} \textbf{Analysis} \\
     \textbf{Output}
   \end{tabular}};

 % edges
 \draw [->] (truth) -- node [above] (dgp)
 {  \begin{tabular}{c} Data-Generating \\ Mechanism \end{tabular}}
 (simdat);

 \draw [->] (simdat) -- node [above] (analysis)
 {  \begin{tabular}{c} Data \\ Analysis \end{tabular}}
 (output);

 \draw [->] (output.south) to [bend left = 20]  node [below, sloped] (performance)
 {   \begin{tabular}{c} Performance \\ Evaluation \end{tabular}}
 (truth.south);

  % nodes related to edges
 \node [rectangle, draw, rounded corners = 0.5em] (methods) [above = 1em of analysis]
 {\footnotesize \begin{tabular}{l}
                  \multicolumn{1}{c}{\textbf{Methods}} \\
           Linear regression \\ Random forest \\ $\hdots$  \end{tabular}};

  \node [rectangle, draw, rounded corners = 0.5em] (params) [above = 1em of dgp]
       {\footnotesize \begin{tabular}{l} \textbf{Parameters} \\
                 Sample size \\ Variance \\ $\hdots$  \end{tabular}};

  \node [rectangle, draw, rounded corners = 0.5em] (measures) [right = 5em of performance]
       {\footnotesize \begin{tabular}{l} \multicolumn{1}{c}{\textbf{Measures}}  \\
                 Bias \\ Mean squared error \\ $\hdots$  \end{tabular}};

   \draw [->] (params.south) to (dgp);
  \draw [->] (methods.south) to (analysis);
  \draw [->] (measures.west) to (performance.east);
\end{tikzpicture}
\caption{Schematic illustration of a simulation study.}
\label{fig:simulationstudy}
\end{figure*}

Simulation studies play a crucial role in methodological research, allowing researchers to systematically test and compare statistical (data analysis) methods under controlled conditions \citep{Hoaglin1975}. In such studies, researchers simulate synthetic datasets based on a known data-generating mechanism (DGM), analyze these data using the methods under investigation, and compare their outputs with the known truth to evaluate method performance (see Figure~\ref{fig:simulationstudy} for an illustration). Here, we focus on the evaluation of reusable methods for tasks such as missing-data imputation and treatment-effect estimation, rather than on ``bespoke model building'', where scientists build highly customized models for a specific use case. These methods are commonly applied across studies in different contexts, making it important to understand their performance across a range of conditions relevant to their intended use. However, even the aforementioned customized models often consist of components that are reused across contexts, such as certain estimators or missing data imputation strategies, whose evaluation is the concern of the framework.

Simulation studies, in contrast to purely theoretical assessment, can reveal practical strengths, limitations, and boundary conditions of statistical methods under complex conditions. However, as in empirical research in other scientific fields, conflicts of interest and misaligned incentives can compromise the validity of simulation studies. Researchers developing new statistical methods are often expected to demonstrate the new method's superior performance in a simulation study, resulting in pressure to demonstrate the new method in the best light \citep{norel2011, Boulesteix2015, Boulesteix2020, Mandl2026}. This pressure is particularly concerning given that simulation studies in methodological research have been shown to suffer from design, execution, and reporting issues almost since their inception \citep{Hoaglin1975, Hauck1984, Koehler2009, Harwell2018, Morris2019, Siepe2024, Williams2024}. 

% Design:
In the \textit{design} of a simulation study, researchers may often rely on ad-hoc or oversimplified parametric DGMs to simulate data \citep{Boulesteix2020B, Sauer2025}. The simulation study introducing a new method may be prone to using DGMs that favor the novel method's performance \citep{Ullmann2022, Niessl2021, breiman2001, Niessl2024}. Small changes to the DGM can produce markedly different outcomes, complicating method evaluations and reducing generalizability \citep{pateras2018, vanSmeden2016, fairchild2024}. Ambiguity in the objectives and evaluation criteria of a simulation study leaves room for flexible execution and reporting choices that might bias the between-method comparison. 

% Execution: 
There are many researchers' degrees of freedom in the \textit{execution} of a simulation study \citep{Pawel2024}. When different methods are compared, researchers may be less familiar with methods that they did not develop themselves \citep{Niessl2024}. This can lead to a suboptimal implementation of competitor methods, again favoring the method of choice. Due to computational issues or non-convergence of a method, simulation repetitions may not always lead to valid outputs \citep{Paweletal2025, Wuensch2025}. Different ways of handling such issues can change the main results and implications of a simulation study. Many other small decisions, such as the choice of hyperparameters, the removal of certain conditions, or a change in the evaluation criteria, can also hinder an impartial evaluation \citep{Pawel2024}.

% Reporting
\textit{Reporting} issues in methodological research are similar to those found in other empirical research \citep{Hoaglin1975, Hauck1984, Harwell2018}. Simulation studies often contain a large number of results, and selective reporting of outcomes can easily favor a specific method \citep{Pawel2024}. Rather than weighing the trade-offs between different methods in different contexts, simulation results are sometimes reported in a way that seeks to find the ``best'' overall method, which is often unrealistic and unhelpful \citep{Strobl2024}. The Monte Carlo uncertainty of performance estimates is rarely reported in simulation studies, which is especially problematic in computationally intensive studies using relatively few simulation repetitions \citep{Morris2019, Siepe2024, Koehler2009}. Several literature reviews have found that code to reproduce a simulation is often not shared and that insufficient information on the computational environment is provided \citep{Siepe2024, Morris2019, Williams2024, Kucharsky2020}. This hinders the reproducibility and replicability of simulation studies, and the reuse of design choices in new studies \citep{Luijken2024, Lohmann2022}.

Overall, these issues lead to a proliferation of simulation studies that introduce, evaluate, and favor novel methods. Consequently, different contributions to statistical methodology result in conflicting conclusions and recommendations \citep{fairchild2024, lang2024}. One might speculate that those discrepancies confuse empirical researchers who have to choose a method for analyzing their data, result in the delayed adoption of effective statistical approaches in practice, and hinder further methodological advancements \citep{lotterhos2022, Boulesteix2025}.

To overcome these issues, we propose to disentangle the development of statistical methods and the design of simulation studies via the adoption of \emph{living synthetic benchmarks}. This proposal is inspired by recent advances in machine learning and artificial intelligence research; standardized benchmarks such as the ImageNet \citep{deng2009imagenet} and Critical Assessment of Structure Prediction \citep[CASP, ][]{moult1995large} competitions have resulted in groundbreaking scientific developments like AlexNet \citep{krizhevsky2012imagenet} and AlphaFold \citep{jumper2021highly}. Central benchmarking repositories, such as the UC Irvine Machine Learning Repository \citep{Kelly2023}, have facilitated methodological research in the machine learning community since 1987, and other benchmarking platforms, such as OpenML \citep{Bischl2025} and Hugging Face (\href{https://huggingface.co/}{huggingface.co}), have emerged. In computational biology, rigorous benchmarking has also been used to navigate the large number of available methods \citep{weber_essential_2019, robinson_benchmarking_2019, Mallona2024a, Mallona2024b}. Such standardized benchmarking platforms provide neutral, transparent, third-party data and performance measures, facilitating objective methodological comparisons and cumulative improvements. Here, we focus on the idea of synthetic benchmarks, by which we mean a standardized and impartial set of simulated data that can be used to compare statistical methods. Hereafter, we refer to this idea as ``benchmarking'' for simplicity. We will outline its relation to classical benchmarking in the discussion section.

Adopting such benchmarking frameworks in methodological research could help overcome current limitations. Standardized benchmarks would provide neutral, reproducible datasets and scenarios for evaluating competing methods. As such, continuous evaluation of new statistical methods would be especially useful for ``late-phase'' and confirmatory methodological research where the aim is to conduct comprehensive and neutral comparisons to guide data analysis decisions \citep{Heinze2024, lange2025}. 

In the following, we outline a blueprint for building and maintaining such benchmarks, and discuss technical and organizational challenges of implementation (Section~\ref{sec:blueprint}). We then illustrate its feasibility with a prototype benchmark of publication bias adjustment methods in meta-analysis (Section~\ref{sec:example}). Finally, we discuss the advantages and limitations of the approach along with concluding recommendations (Section~\ref{sec:discussion}).

\section{Blueprint of Synthetic Benchmarking}
\label{sec:blueprint}

\begin{figure*}[h]
    \includegraphics[width=\textwidth]{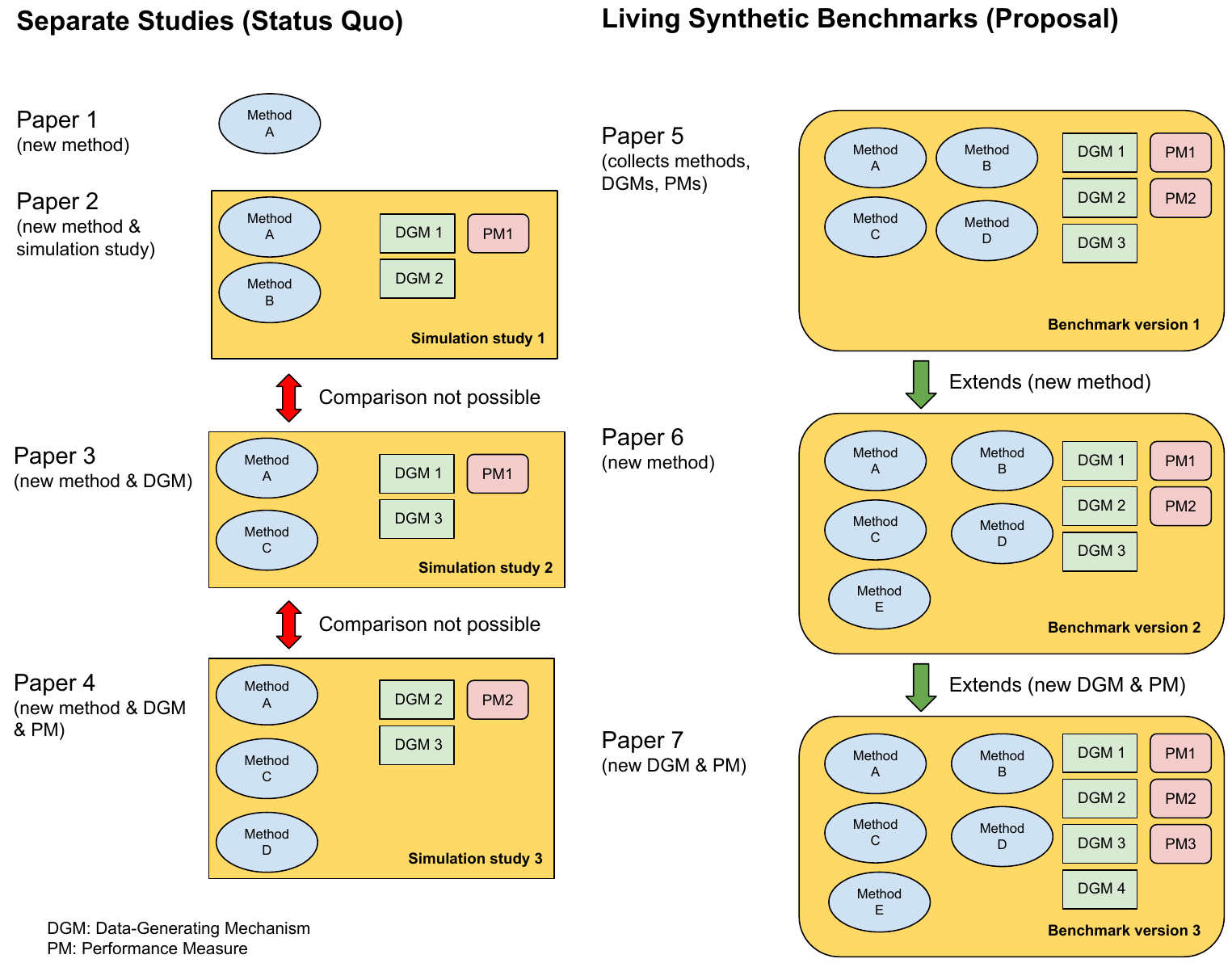}
    \caption{Comparison of standard methodological development process via separate studies with the proposed methodological development process via living synthetic benchmarks}
    \label{fig:workflow}
\end{figure*}

% introduce DGM, performance measures, simulation, methods within the first sentences
% disentangle DGM, method, performance measures visually. Put "simulation" at a higher level
% think about using ven diagrams for non-overlapping dgm/performance measures/simulation methods
The current state of the methodological literature often consists of incompatible and isolated simulation studies. The left panel of Figure~\ref{fig:workflow} offers a simplified depiction of this status quo. The main element is a simulation study (large yellow square), which compares several statistical methods (Method; e.g., linear regression) on a set of performance measures (PM; e.g., mean squared error) using a set of data-generating mechanisms (DGM; e.g., artificial data generated from a regression model).

Initially, Paper 1 introduces a new Method A to aid in answering a substantial question. Then, Paper 2 introduces Method B and compares it with Method A using PM1 and DGMs 1 and 2 in Simulation study 1. The first emerging issue is that Paper 2 may have designed, executed, and reported Simulation study 1 in a way that benefits the newly proposed Method B (e.g., selection of PMs and DGMs). Later, Paper 3 introduces Method C and compares it with Method A in a new Simulation study 2. Paper 3 may have the same issues as Paper 2. In addition, the newly introduced Simulation study 2 misses Method B and DGM 2, which makes the results of Simulation study 2 not fully comparable to the results of Simulation study 1. Finally, Paper 4 and subsequent papers may continue simultaneously introducing new methods, PMs, and DGMs, which complicates between-method and between-simulation comparisons.
The right panel of Figure~\ref{fig:workflow} visualizes the proposed changes to the method development process: introduction of living synthetic benchmarks. In summary, the proposal consists of a) collecting existing methods, DGMs, and PMs to create the initial benchmark, b) separating the subsequent introduction of new methods, DGMs, and PMs, and c) ensuring cumulative and continuous reporting. The following subsections detail the individual parts.

\subsection{Creating the Initial Benchmark}
% once the field reaches some level of maturity, the continuous benchmarking can begin
% similar to performing a literature review: overviewing what simulations were already employed and what were they critical compomnents -> combinging them into crucial DGMs
% collecting all the relevant methods to be applied to those DGMs
% you could have DGM-only papers, scrutinizing different DGMs
% separating data simulation and data analysis team: Kreutz et al., 2020
% adversarial collaboration between researchers: Binder et al., 2020 

% the field reaches some level of maturity: i.e., existence of methods and simulations
The goal of this phase is to combine previously developed methods, DGMs, and PMs into an initial benchmark. Benchmarking cannot begin from scratch; the methodological subfield must attain some level of maturity by establishing an initial set of methods and simulation studies. This body of previous literature allows for establishing issues the subfield is trying to address, along with a variety of opinions and approaches. Although there is no predetermined number of DGMs, PMs, and methods needed to warrant a benchmark, having at least two methods, one DGM, and one PM constitutes a bare minimum.

% summarizing and reviewing the existing literature
The first step of establishing the benchmark is reviewing the literature and creating an overview of already published methods, DGMs, and PMs, akin to a systematic review but focused on methodological research \citep[for some examples, see e.g.,][]{Langan2016, Zhang2017, Hinds2018, Smid2019, Morel2022, Abell2023}. While individual methods usually form distinct entities, new simulation studies often partially build upon previous simulation studies. As such, it is important to clearly define the common and distinct features of the existing simulation studies to prevent duplicating the same methods, DGMs, and PMs in the benchmark.

Subsequently, the different DGMs should be examined in depth to verify that they represent the intended assumptions. This will usually entail that the simulated data are comparable to real-world applications, but some benchmarks may also focus on testing extreme cases or (near-)asymptotic behavior. Although simulations alone cannot establish how closely a DGM approximates reality, evaluating methods across a diverse range of DGMs and parameter settings can reveal strengths and limitations that would remain hidden in a narrower simulation study. The value of a benchmark therefore depends not only on the realism of individual DGMs, but also on the breadth and diversity of the overall simulation design within the methods' intended scope. Often, existing simulations will need to be modified to match the desired properties. Furthermore, a complete re-implementation and replication of published simulation studies might be needed due to the missing or incompatible code and data. This is a positive side effect of our proposal, aligned with the recent calls for replications of simulation studies \citep{Lohmann2022, Luijken2024}.

% finalizing the initial benchmark
Once all relevant DGMs, methods, and PMs are identified, the benchmarking infrastructure needs to be set up (see Section~\ref{sec:technical-implementation}). The methods need to be evaluated with all PMs across all DGMs, and the initial benchmark results should be published.

% not sure where to put this
Rather than one research group initiating a benchmark, an alternative approach might be to create a special issue that allows multiple authors and groups to submit DGMs, PMs, and methods. For example, the initial benchmark establishment could be an ``adversarial collaboration'' process \citep{Cowan2020} that ensures that no researcher can skew the initial benchmark. Similarly, the benchmark can be created in a distributed process, with some groups setting up the DGMs while other groups implement the methods \citep[as in][]{Kreutz2020, Perperoglou2025}.

\subsection{Addition of Methods}

Once the initial benchmark is established, researchers developing new methods ought to evaluate the performance of these methods using the benchmark. Apart from disentangling the conflicting incentives of simultaneous methods, DGM, and PM development, using the existing benchmark greatly simplifies the contribution of new methods. First, researchers who develop a new method do not need to develop new DGMs simultaneously. Instead, they import the existing benchmark and evaluate the new method using already-designed DGMs and PMs. Second, researchers do not need to recompute the results for the other methods. Instead, they use the precomputed results already present in the benchmark. This becomes especially helpful as some methods may require a large amount of computational resources and/or expertise (e.g., in implementation or selecting hyperparameters). Once the results of the new method are obtained, the researchers ought to upload the results to the existing benchmark so future researchers can benefit from the already computed results.

One complicating matter could be different opinions on what constitutes a ``method'' \citep{Morris2019, Paweletal2025}. This term can refer to a more abstract notion of a data analysis procedure (e.g., random forest), its implementation (e.g., the \texttt{ranger} random forest implementation from \citealp{Wright2017}), or the combination of these with a specific hyperparameter setting (e.g., the \texttt{ranger} random forest implementation with hyperparameter $\mathtt{mtry}=\sqrt{p}$ where $p$ is the number of predictors). The more fine-grained such a definition is, the larger the database may become over time, which may lead to computational and interpretation issues. Specifying ``method'' and ``version'' designations seems to be a minimum requirement that allows for updating (and deprecating) of existing methods; however, there is no one-size-fits-all solution to this issue.

\subsection{Addition of Data-Generating Mechanisms and Performance Measures}
% separate addition of DGM puts spotlight on high quality data generation
Establishing benchmarks also emphasizes the importance of high-quality DGMs. Rather than focusing solely on developing new methods, researchers can contribute relevant DGMs and PMs. These contributions should be reviewed with the same scrutiny as any methodological contribution. This will ultimately produce more meaningful and realistic method evaluations compared to the status quo of DGMs that are often very simplistic since they are often merely an afterthought of method development \citep{Boulesteix2020B, Sauer2025}.

Once new DGMs and PMs are introduced, researchers ought to compute the performance of the included methods on these new settings and update the benchmark with these results. As such, future researchers again benefit from the already computed results.

% not sure where to put this
The development of new DGMs can also provide new challenges for the subfield. For instance, researchers might identify previously unexplored settings, such as specific types of missing data or violations of distributional assumptions, and create DGMs that closely mirror the assumed mechanism. A potentially lacking performance of existing methods under the new settings would highlight the need for further methodological improvement. Systematically varying assumptions across DGMs also provides a form of sensitivity analysis, showing how method performance changes under departures from its motivating model. Living benchmarks make these assessments reusable and cumulative as new methods and DGMs are added.

\subsection{Cumulative and Continuous Reporting}
% new performance measures?
% true "confirmatory" research possible (Lange et al., 2025)

The continuous integration of methods, DGMs, and PMs while updating the benchmark allows for cumulative and sequential reporting on an openly accessible platform (e.g., a website). Consequently, practitioners in need of guidance on method choice do not need to disentangle the ever-growing jungle of methodological articles that introduce new methods and simulation studies. Instead, they can consult comprehensive and more impartial benchmarking results to identify suitable methods for their specific needs. It is usually impossible to identify a single overall winner, but the benchmark might help in understanding how the method's performance depends on data characteristics \citep{Strobl2024}. A method can be valuable if it performs adequately across a range of relevant settings, even if it does not offer the best performance in any single one of them. An important goal of a benchmark is therefore not to crown a best method, but to rule out clearly poor ones and to provide researchers with information on how to choose between the remaining ones.

Benchmark results can further guide method choice by distinguishing observable data characteristics from aspects of the DGM that remain unknown in an application \citep{Sauer2025, Strobl2024}. For example, the sample size and proportion of missing observations are observable and can help practitioners identify relevant benchmark settings for comparing imputation methods. The mechanism responsible for missingness, however, usually remains uncertain. Within settings with similar observable characteristics, comparisons across plausible missingness mechanisms can reveal whether a method maintains adequate performance or depends strongly on particular assumptions. A benchmark can therefore support both recommendations based on observable features and an assessment of performance across aspects of the DGM that remain unknown.

While this process is straightforward in theory, navigating potentially hundreds of DGMs, methods, and PMs can be challenging in practice. For descriptive purposes, aggregated PMs and a ``leaderboard'' ranking methods by aggregated performance may help summarize these results. For example, it might be useful to report the minimum, maximum, mean, and median PMs of methods across a set of DGMs. However, it is crucial to ensure that the selected PMs can be meaningfully aggregated across different DGMs. For instance, mean squared error and bias (and other scale-dependent PMs) of regression coefficients may not be comparable across DGMs with different scales (e.g., one DGM simulates data in kilograms and another in pounds). A possible solution is to use ranks or relative performance scores when aggregating scale-dependent PMs or to restrict the aggregation within the same class of DGMs. Another potential issue with aggregation is the risk of ``gaming'' the overall leaderboard by overfitting methods to perform well on aggregated scores \citep{Singh2025}. A choice must also be made whether to average ranks across all DGM conditions or across different DGM sets. For example, a DGM set could be all DGM conditions from a previously published simulation study. If different DGM sets contribute a varying number of DGM conditions, averaging across DGM sets risks distorting the aggregate scores by introducing redundant DGM sets into a simulation (i.e., it is easier to shift the overall rank by adding new DGM sets). However, averaging across DGM conditions risks distorting the aggregate score by introducing redundant DGM conditions in a new DGM set (i.e., it is easier to shift the overall rank by adding a DGM set with many conditions). Importantly, all aggregation and ranking must be carefully communicated in order to avoid misinterpretations, such as interpreting ``best aggregated performance'' as meaning a method is the ``best'' method for every application \citep{Strobl2024}. 

Data processing choices, such as the handling of non-convergence or other missingness, can affect the benchmark results in numerous ways. For example, if a method does not converge in very challenging simulated datasets (e.g., with small sample sizes), excluding the missing repetitions from that method may bias performance in its favor. It is therefore helpful to also report convergence of methods and provide results for different missingness handling strategies \citep{Paweletal2025, Wuensch2025}. In many cases, repetition-wise deletion (i.e., removal of all repetitions with at least one missing method) might not be a feasible strategy since the addition of new methods with different patterns of missingness could impact the previously reported results of the existing methods and eventually lead to the removal of all repetitions as the number of methods increases. Method-wise deletion (i.e., removal of all repetitions with missing results on a method-by-method basis) prevents this issue; however, it might result in potentially incomparable results between methods, as the convergence is tied to the ``difficulty'' of the repetition (see the example in \citealp{Paweletal2025}). Another alternative is method replacement (i.e., replacing missing values with a simpler version or a default method), which can mimic ``analytic pipelines'' applied in practice. For example, if a complex regression method does not converge, an analyst would not give up but apply a simpler linear regression. While the method replacement might be intriguing when communicating results and recommendations to practitioners, it obscures the actual performance of any single method. It is therefore important to consider the intended audience when communicating missingness handling strategies, and to provide PMs for different strategies to assess their robustness.

% The disentanglement of methods, DGMs, and PMs introduction also produces an impartial comparison environment, corresponding to the true late-stage of confirmatory methodological research \citep{Heinze2024, lange025}. 

\subsection{Retiring Deprecated Benchmark Components}

Over time, some of the DGMs, methods, and PMs may become deprecated or practically irrelevant. If in such a case there is a consensus in the field that a benchmark component is outdated, it may be removed from the benchmark to simplify future reporting and save computing resources. How exactly such a consensus should be reached (e.g., by voting) strongly depends on how a benchmark is organized. This is still an open question, which we will discuss further in Section~\ref{sec:discussion}.

\subsection{Technical Implementation}
\label{sec:technical-implementation}

While the details of the technical implementation will vary from subfield to subfield, we believe that successful benchmarking has several technical prerequisites. First, DGMs must be open and reproducible so that anyone developing a new method can download, run, and extend the benchmark with their results. In many cases, it might be feasible to share the raw simulated data, which dismantles any fears of cross-platform reproducibility (e.g., differing random number generators across operating systems) and versioning (e.g., differing random number generators across different versions of the software).
Second, the raw method-simulation-repetition level results must be open and reproducible. As such, anyone can verify the submitted PMs or compute new PMs. %Ideally, there would be a list of dedicated reviewers handling new benchmark submissions. 
Third, the benchmark should follow the FAIR data sharing guidelines as closely as possible, which ensure that all shared data are Findable, Accessible, Interoperable, and Reusable \citep{Wilkinson2016}. This implies certain quality standards such as consistent variable naming, metadata, standardized file formats, and persistent identifiers.
% Requirements

% Licensing 
Beyond these general recommendations, there are a number of additional steps that could be taken to ensure the reproducibility and robustness of benchmarking results. These decisions commonly involve a trade-off between greater rigor and reproducibility on the one hand, and flexibility and accessibility on the other. 

% Docker containers? Method version?
% how about different programming languages?
% proprietary software: you can even compare results to proprietary software if you don't have it, because it has been done before. 
In some fields, proprietary software is widely used (e.g., M\textit{Plus} in structural equation modeling, \citealp{muthen2017mplus}, or SAS in pharmaceutical research, \citealp{SAS2025}). While it would be possible to only include open-source software in the benchmark, this may hinder the comparison with popular methods. In such cases, authors of newly submitted DGMs may only be required to evaluate all openly available methods, while other authors with access to proprietary software could extend the benchmark later. The potential to compare methods to proprietary competitors, even when an author does not have access to the software, constitutes an additional advantage of our proposal. 

Achieving full computational reproducibility of all computations should be striven for, but it is not an easy task. It is complicated by the potential use of different programming languages and computing environments. Ideally, each benchmark run would be encapsulated in a software container, as, for example, implemented in the Omnibenchmark system in computational biology \citep{Mallona2024b}. However, such containerized benchmarking is hard to scale \citep{Bischl2025}. Furthermore, in our experience, most statisticians and methodologists performing simulation studies lack training in containerization and other more advanced computational reproducibility approaches. Thus, requiring full containerization could severely limit the adoption of benchmarking by the community. In sum, there is a delicate trade-off between achieving computational reproducibility and accessibility of the benchmark. Nevertheless, as a bare minimum, versions of all software programs and packages, as well as information on the computational environment, should be provided when submitting new results to the benchmark. Even when full computational reproducibility is not achievable, a publicly shared and maintained living benchmark provides a meaningful improvement over the current status quo.

\section{Example: Publication Bias Adjustment Methods}
\label{sec:example}

We now illustrate a prototype benchmark for evaluating meta-analytic publication bias adjustment methods. Publication bias occurs when studies with statistically significant or favorable results are more likely to be published \citep{rosenthal1964further}. As such, it distorts the evidence in the published literature and presents a grave threat to cumulative science in all fields \citep[e.g.,][]{de2015surge, fanelli2017meta, bartos022footprint}, warranting methods that can adjust the available evidence for potential publication bias. We chose this field as it is relatively mature, with the first methods being published more than 40 years ago \citep[for an overview see][]{marks2020historical}. Initially, there were many scattered and incommensurable simulation studies, but some recent large-scale simulation studies attempted to consolidate the findings and recommendations \citep{carter2019correcting, hong2021using}. In Figure~\ref{fig:history}, we overview major simulation studies on publication bias adjustment methods. In total, these have accumulated over 26,000 citations at the time of writing, potentially impacting many meta-analysis decisions in practice. 

\subsection{History of Publication Bias Adjustment Methods}

\begin{figure*}
    \centering
    \includegraphics[width=\textwidth]{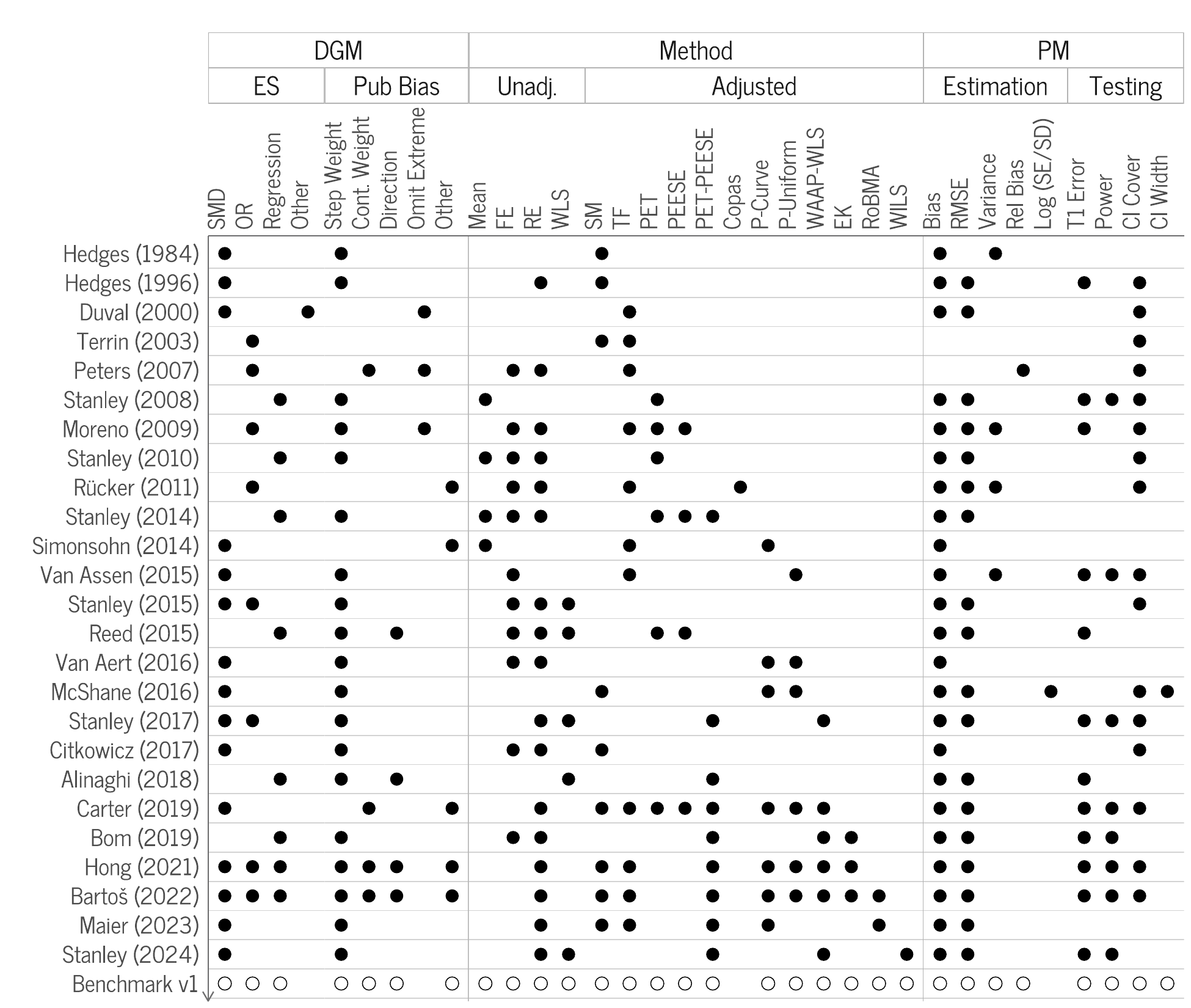}
    \caption{Timeline of major simulation studies on publication bias adjustment}
    \label{fig:history}
    \tiny
    \textit{Note.} DGM = data-generating mechanism, ES = type of effect size, Pub Bias = type of publication bias, Unadj = unadjusted methods, Adjusted = adjusted methods, PM = performance measures, OR = odds ratios, SMD = standardized mean differences, Regression = linear regression, Step Weight = step weight function, Cont Weight = continuous weight function, Direction = selection on the direction of the effect, Omit Extreme = omission of the most extreme results, Mean = unweighted mean, FE = fixed effect, RE = random effects, WLS = weighted least squares, SM = selection model, TF = trim and fill, PET = precision effect test, PEESE = precision effect estimate with standard errors, PET-PEESE = precision effect test and precision effect estimate with standard errors, COPAS = Copas selection model, WAAP-WLS = weighted average of adequately powered studies with weighted least squares, EK = endogenous kink, RoBMA = robust Bayesian meta-analysis, WILS = weighted and iterative least squares, RMSE = root mean squared error, Rel Bias = relative bias, Log (SE/SD) = logarithm of the estimated standard error of the estimates divided by the standard deviation of the estimates, CI = confidence interval, T1 Error = Type I error.
    For Peters (2007) \cite{peters2007performance} and Stanley (2015) \cite{stanley2015neither}, the specifics of the selection functions used in the DGM were not fully clear to us, but we nevertheless coded them as present. 
\end{figure*}

Publication bias adjustment methods such as selection models and trim and fill were developed early on \citep[e.g.,][]{hedges1984estimation, duval2000nonparametric}. While some of the subsequent simulation studies compared the main approaches at the time \citep[e.g.,][]{terrin2003adjusting}, other influential simulation studies examined only a limited subset of methods \citep[e.g.,][]{peters2007performance, moreno2009assessment} that were introduced later (e.g., subset and regression approaches). New articles then introduced new methods (PET, PET-PEESE, p-curve, p-uniform, endogenous kink), DGMs, and PMs while not including all of the relevant approaches \citep[e.g.,][]{stanley2014meta, simonsohn2014pcurve, van2015meta, mcshane2016adjusting, alinaghi2018meta, stanley2017finding, bom2019kinked}. Importantly, the limited overlap in DGMs further limited the between-method comparison, as is visible in Figure~\ref{fig:history}, where, for many rows (studies), only a few points (methods, DGMs, PMs) are available. A rare exception was Carter and colleagues \cite{carter2019correcting}, who included a comprehensive comparison of many existing methods, and in contrast to the majority of papers, did not introduce new methods.

A major change came with the article by Hong and Reed \cite{hong2021using}, which was one of the inspirations for our proposal. These authors collected four previously published simulation studies: \cite{stanley2017finding, alinaghi2018meta, bom2019kinked, carter2019correcting} and performed a comprehensive comparison of the existing methods. They openly shared the simulation data and materials, which allowed subsequent authors developing new methodology, \citealp[e.g.,][]{Bartos2023}, to evaluate their methods on an existing set of DGMs.

\subsection{Benchmark Prototype}
  
Hong and Reed \cite{hong2021using} and Bartoš and colleagues \cite{Bartos2023}, the most comprehensive simulation studies on publication bias adjustment to date (see Figure~\ref{fig:history}), could be considered an example of benchmark development, with Bartoš and colleagues \cite{Bartos2023} an example of a method addition to the existing benchmark. While both author teams shared simulation code and results, their formatting does not allow for an easy extension with new methods, DGMs, and PMs, and thus falls short of the proposed benchmarking methodology. Therefore, we collected and reformatted the materials from Hong and Reed \cite{hong2021using} and Bartoš and colleagues \cite{Bartos2023} and replicated the simulation study to simplify future extensions and reuse in the spirit of continuously updated living benchmark. 

\begin{figure*}
    \centering
    \includegraphics[width=0.48\linewidth]{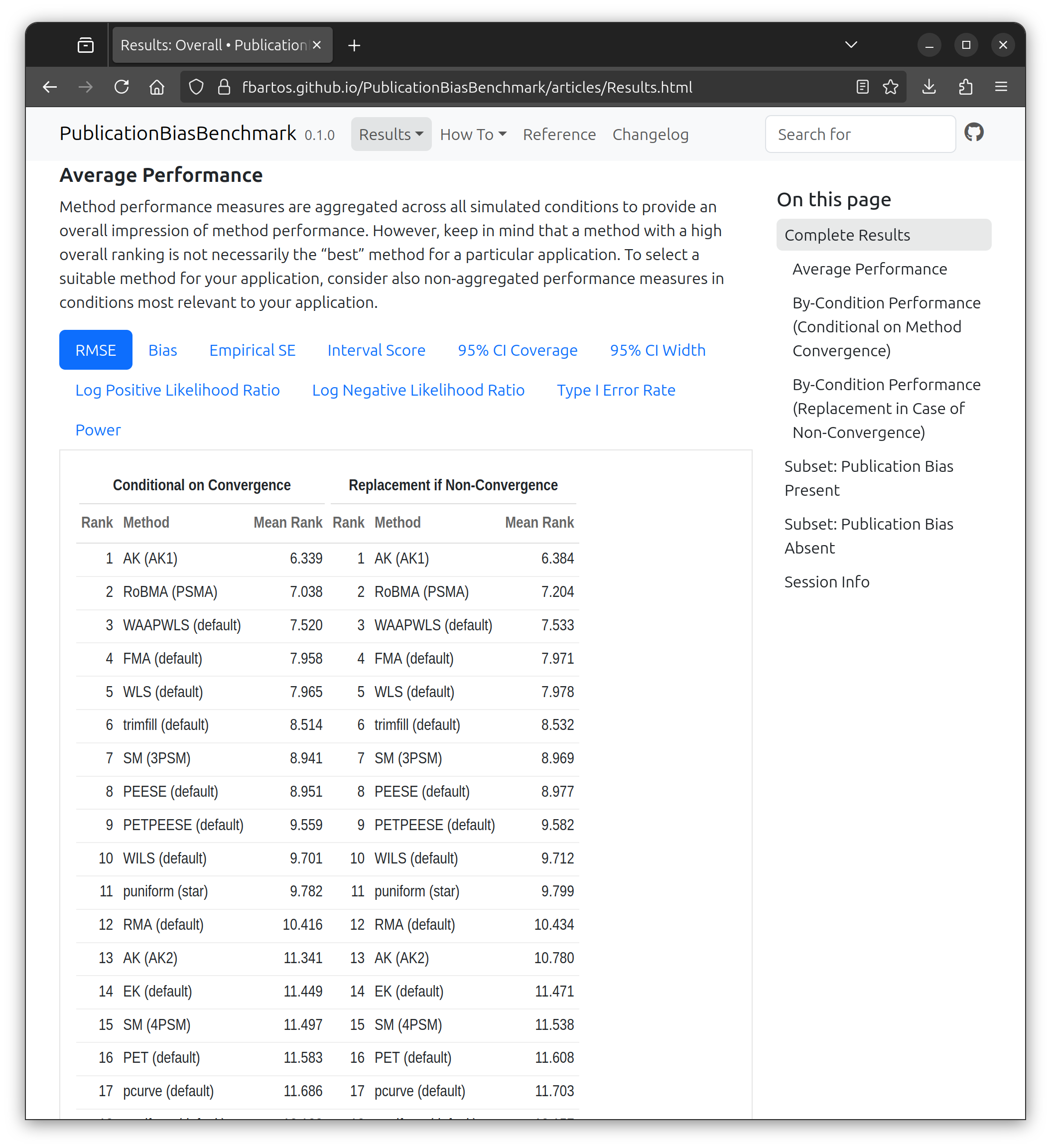} 
    ~ \hfill ~
    \includegraphics[width=0.48\linewidth]{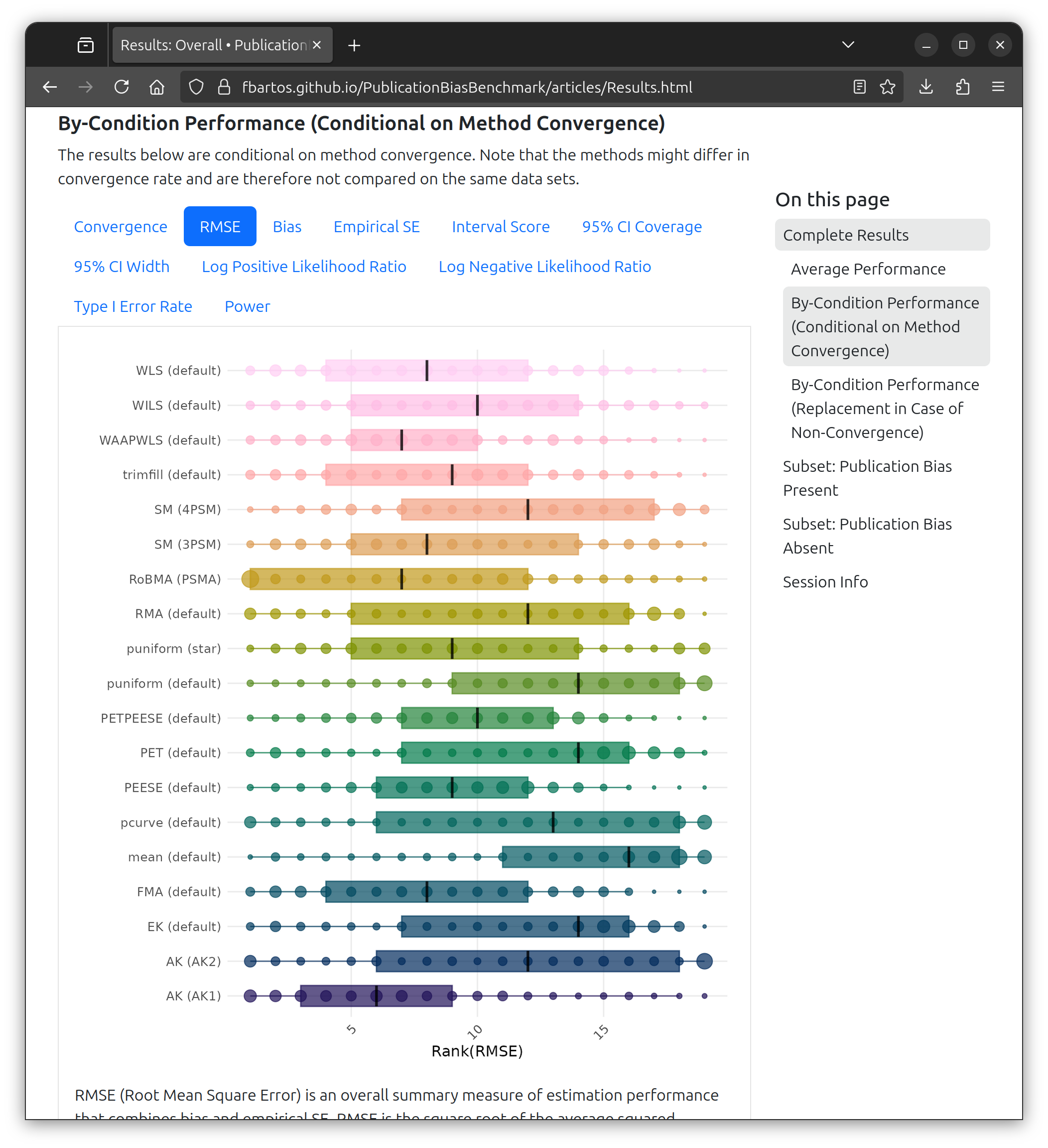} 
    \caption{Screenshots of the interactive benchmark results website (\href{https://fbartos.github.io/PublicationBiasBenchmark/}{fbartos.github.io/PublicationBiasBenchmark/}) showing the ranking of RMSE across all DGMs in table form (left) and figure form (right).}
    \label{fig:webpage}
\end{figure*}

Since almost all contemporary publication bias adjustment methods are implemented in the programming language \texttt{R}, we implemented the new benchmark prototype in a new \texttt{R} package named \texttt{PublicationBiasBenchmark}. The package facilitates reuse, extension, and evaluation of publication bias adjustment methods. Its source code is hosted on GitHub (\href{https://github.com/FBartos/PublicationBiasBenchmark}{github.com/FBartos/PublicationBiasBenchmark}), and it is connected to an Open Science Framework (OSF) repository (\href{https://osf.io/exf3m/}{osf.io/exf3m/}) that stores the simulated data, computed results, and PMs. The package automatically generates a website that summarizes the main results (\href{https://fbartos.github.io/PublicationBiasBenchmark}{fbartos.github.io/PublicationBiasBenchmark}); see  Figure~\ref{fig:webpage} for an example showing the ranking of methods with respect to RMSE (root mean squared error) across all DGMs. Below, we describe the implementation in further detail.

\subsection{Implementation}

First, the benchmark contains the implementation of the four DGM sets included in Hong and Reed \cite{hong2021using} with a standardized data output format and pre-simulated datasets hosted on the OSF repository. The data generation is documented with computationally reproducible scripts. Since data are stored as .csv files, they can be reused across programming languages and statistical software. The package further features functions that allow convenient download and reuse of the pre-simulated data directly from \texttt{R}.

Second, the benchmark implements the methods compared in Hong and Reed \cite{hong2021using} with additional considerations for extensions and output standardization. We further extended the benchmark with two newly developed publication bias adjustment methods: WILS \citep[weighted iterative least squares][]{stanley2024harnessing} and p-uniform* \citep{vanaert2025puniform}. We also updated the implementation of the included methods to follow current best practices. For example, random-effects meta-analysis methods now use the Knapp-Hartung adjustment \citep{knapp2003improved} and a multilevel component with clustered standard errors (in the presence of clustered effect sizes; \citealp{konstantopoulos2011fixed, tipton2015smallsample}) by default (see the package documentation for implementation details of each method). To guarantee full reproducibility, we re-computed results for all methods based on the re-simulated datasets. As with the datasets, the methods' results are stored in a standardized format as .csv files on OSF with the corresponding reproducible scripts. The package features functions for the convenient downloading and reuse of the results from \texttt{R}.

Third, the benchmark contains PMs based on the re-computed results. As such, anyone wishing to extend the benchmark with a new method can simply run their new method on the pre-simulated data and submit the method results (in the same standardized format as the existing methods, see the package vignettes for detailed guidance). The benchmark then contains scripts to automatically update the PMs and the webpage based on the existing results. In addition, we implemented additional PMs not included in the previous simulation studies, such as the interval score (which evaluates CI coverage and CI width simultaneously; \citealp{winkler1972decision}) and positive and negative likelihood ratios (which evaluate Type I error rate and power simultaneously; \citealp{huang2023relative, pepe2003statistical}). For PMs that we implemented in the same way as Hong and Reed \cite{hong2021using}, e.g., Type I error rate, we were able to closely replicate their results with some minor differences due to Monte Carlo error. Note that some PMs (e.g., bias and RMSE) are not directly comparable with the previously reported results, as Hong and Reed \cite{hong2021using} winsorized the results and averaged scale-dependent PMs across DGMs with different scales, which is not done in our benchmark.

\subsection{Results}

The compiled benchmark results enable several novel insights (see Figure~\ref{fig:webpage} for an illustration of pre-computed summaries, with detailed results accessible on the benchmark website and \texttt{R} package). First, examining the publication bias unadjusted methods, we find that fixed-effect meta-analysis and weighted least squares (WLS; both of which were not included in Hong and Reed \cite{hong2021using}) perform, on average, better in estimation-focused PMs (e.g., RMSE and bias) than random-effects meta-analysis. This is unsurprising as they place less weight on smaller studies that are generally affected by publication bias to a greater degree. Interestingly, WLS showed, on average, slightly better CI performance than random-effects meta-analysis, although both methods exhibited severe CI under-coverage across the DGMs.

Second, one of the two newly included publication bias adjustment methods shows somewhat unexpected behavior. While p-uniform* seems to be an improvement upon the previous version of p-uniform in both the Stanley and Alinaghi DGM sets, consistently performing alongside the other selection model-based methods, it shows somewhat erratic behavior in the Bom and Carter DGM sets. This behavior seems to result from a low number of (or a complete lack of) statistically non-significant results in some of the repetitions which negatively impacts the weight function estimation, an issue previously noted by the original developers \citep{vanaert2025puniform}. WILS showed an average performance based on the estimation PMs; however, it lagged behind in terms of confidence interval PMs in all DGMs.

Third, the newly introduced PMs allow us to gain additional insights. For instance, positive likelihood ratios, which compare the probability of obtaining a statistically significant test result under the alternative versus the null hypothesis, are, on average, low for almost all methods, with RoBMA being the only exception. This means that statistically significant tests of their meta-analytic effect carry little information about the presence of the effect for almost all methods (the methods are almost as likely to result in a statistically significant test under the alternative as under the null hypotheses). This is a consequence of the very high Type I error rate those methods exhibit across DGMs. Conversely, the methods generally show a relatively good negative likelihood ratio. As such, the statistically non-significant tests of the meta-analytic effect are relatively informative with respect to its absence.

To sum up, the newly developed publication bias adjustment method benchmark allows us to obtain new insights into methods, DGMs, and PMs that were not assessed simultaneously before. The \texttt{R} package implementation is designed to allow for easy extensions with new DGMs and methods as documented in the accompanying vignettes. This should not be regarded as the final benchmark for publication bias adjustment. The included DGMs should be thoroughly scrutinized, and additional DGMs should be incorporated to reflect recent developments in the field, such as methods for handling dependent effect sizes and meta-regression. Together, these steps will form a natural evolution of the benchmark. As such, it will facilitate future research on publication bias adjustment methods. Moreover, researchers from other fields can re-purpose the package by replacing the existing DGM, method, and PM objects, and by applying the main package structure to different settings.

\section{Discussion}
\label{sec:discussion}
%In this article, we proposed a novel way to compare statistical methods via standardized synthetic benchmarks. Instead of disparate simulation studies investigating a small number of settings and methods mainly proposed by the authors, these benchmarks offer wide-ranging and impartial comparisons between different methods. We now summarize how this proposal relates to the status quo in methodological research, discuss logistics of implementation, and provide final recommendations for the uptake of benchmarks.  

% Some points on status quo
Contemporary methodological research suffers from similar issues as psychological measurement in empirical research \citep[see e.g.,][]{flake2020measurement}. There is a strong incentive to publish new measures/methods; novel measures/methods are mostly evaluated by those proposing them, and these evaluations are often unstandardized and superficial. Thus, rephrasing the words of Elson and colleagues \cite[p.1]{elson2023psychological}: ``\emph{[Simulation studies] suffer from the toothbrush problem: no self-respecting analyst wants to use anyone else's.}'' Although authors often describe their simulation studies as ``extensive'', many function mainly as proof-of-concept demonstrations rather than rigorous evaluations across realistic DGMs and competitor methods. This is generally understood within the methodological research community but may be less clear to potential method users, who often rely on evidence from simulation studies to choose appropriate methods. To address this, we advocate a clearer separation between proof-of-concept and rigorous simulation studies, a goal that could be advanced through the use of living synthetic benchmarks.

We do not argue for the complete elimination of simulation studies in manuscripts that present new methods. Such studies represent an ``early-phase'' of methodological research \citep{Heinze2024} and allow methodologists to verify the basic performance of a newly proposed method, as well as identify conditions under which the method fails. However, methodologists should also consider evaluating their methods on living synthetic benchmarks if available, in order to compare their method directly with existing alternatives. A neutral evaluation of this kind would enable stronger claims about performance in the spirit of ``late-phase'' methodological research \citep{Heinze2024}. If those benchmarks do not exist yet, methodologists should consider establishing them to facilitate neutral and cumulative method development.  

Evaluating methods across multiple DGMs closely relates to sensitivity analyses in simulation studies, where assumptions and parameter settings are varied to examine how the findings depend on the chosen scenarios and find conditions where methods ``break'' \citep{Morris2019}. Such evaluations help establish whether conclusions about a method's strengths and limitations persist across alternative settings. Living synthetic benchmarks make these simulation-based assessments reusable and cumulative, allowing new methods to be evaluated under previously investigated settings and existing methods to be evaluated under newly introduced settings.

Synthetic benchmarks should neither replace nor diminish real-data applications or ``classical'' benchmarks on real data. In methodological research, applications demonstrate how a method addresses a scientific question, how it is applied, and how its results are interpreted \citep{Heinze2024}. They can also reveal practical difficulties and model inadequacies that motivate further investigation. However, an application does not by itself establish that the assumed model closely approximates the true DGM or that the method is robust to departures from that model. There is also a risk of the ``storytelling fallacy'', where a post-hoc explanation built on a single data set is presented as solid evidence that the method is adequate \citep{Mandl2026}.

Real-data benchmarks for supervised prediction allow methods to be compared by evaluating their predictions against observed outcomes on held-out data \citep{Friedrich2023}. This does not require knowledge of the full DGM. For many inferential tasks, however, the relevant truth is not observed, so properties such as estimation bias and confidence interval coverage cannot generally be assessed from a single real-data analysis \citep{Morris2019}.

There are tradeoffs between synthetic and real data benchmarks \citep{Friedrich2023}. Living synthetic benchmarks allow performance to be evaluated repeatedly under known DGMs, but their conclusions remain conditional on the chosen scenarios. Real-data applications can inform benchmark development by suggesting relevant data structures, parameter settings, and practical problems not yet represented in the included DGMs \citep{Sauer2025}. These can motivate new simulation settings, connecting the continued development of the benchmark to problems encountered in practice.

The presented benchmark prototype is an initial step toward advancing the publication bias adjustment methodology. However, this is only the beginning of a larger effort to establish a more systematic method development process. The current benchmark is limited by the constraints of the previously published DGMs, and future work should critically assess and expand this aspect of the benchmark. Nevertheless, this benchmark can also serve as a prototype for organizing new synthetic benchmarks. Further efforts are required to implement computational templates that enable communities to easily initiate new synthetic benchmarks. For instance, an \texttt{R} package could be developed to automatically generate the necessary folders, files, and functions for compiling results, archiving data on platforms like OSF, and integrating findings into accessible online repositories. Benchmarking systems from other fields, such as Omnibenchmark \citep{Mallona2024b} or OpenML \citep{Bischl2025}, could also potentially be repurposed.

Many computational and logistical questions remain in the adoption of living synthetic benchmarks. For example, some areas of methodological research, such as deep learning or Bayesian statistics, can involve computationally intensive methods, making it burdensome to reproduce simulations across all benchmarking datasets. In such a case, a random subsample of repetitions from a condition may be used to obtain a rough estimate of the method's performance. The continuous nature of the proposed approach alleviates the computational burden from all future comparisons, as the benchmark does not need to be recomputed every time a new method is added. Moreover, journals, societies, or other communities could dedicate resources to compute and maintain such benchmarks, as they would be highly valuable for advancing the field. With an existing benchmark, evaluating a new method becomes easier, especially during peer review. Relatedly, in cases with a large number of DGMs, researchers could define subsets of relevant DGMs for specific research questions, reducing complexity and preventing resource waste. For example, a subset of DGMs could focus on missing data and violations of normality assumptions, enabling practitioners to query the benchmarks more efficiently and obtain the most relevant model comparisons for their specific problems.

Issues around computational burden are also tied to questions of centralization versus decentralization in the benchmarking infrastructure. A more centralized system would have designated organizers responsible for creating the benchmark, maintaining the computational infrastructure, and ensuring the validity, reproducibility, and interoperability of new submissions. While this approach has the advantage of potentially higher quality standards and prevents fabricated submissions, it introduces dependency on a few maintainers, complicates benchmark expansion, and, crucially, raises concerns about impartiality. A potential solution is to host the benchmark on a platform such as GitHub, where researchers either open pull requests with novel additions to the benchmark and maintainers check the reproducibility on a random subset of results, or fork the existing benchmark in cases where the existing maintainers cease to support its continuation.

In conclusion, the long-term progress of methodological research depends on moving beyond isolated simulation studies toward cumulative and standardized evaluation. Instead of creating a new simulation study for each proposed method, researchers should apply their work to existing benchmarks. New simulation studies, representing emerging challenges in the field, could be added to these benchmarks, and existing methods can then be reapplied for comparison. Systematic reviews and meta-analyses can play an important role by collecting DGMs and method implementations into a living, continuous benchmarking system \citep{Langan2016, Zhang2017, Smid2019, Morel2022, Abell2023}. Ultimately, such coordinated efforts would transform methodological research into a cumulative enterprise, one in which progress can be more transparently demonstrated and more credibly assessed.

\section*{Disclosure Statement}
František Bartoš declares himself to be an author of Robust Bayesian Meta-Analysis (RoBMA), one of the methods compared in the benchmark. All of the DGMs, however, were developed by other researchers.

\bibliographystyle{icml2024}
\bibliography{bibliography}

\end{document}